\documentclass[aps,prl,superscriptaddress,twocolumn,longbibliography]{revtex4}
\usepackage{amssymb}
\usepackage{graphicx}
\usepackage{amsmath}
\usepackage{MnSymbol}
\usepackage[colorlinks=true,linkcolor=blue,citecolor=blue,urlcolor=blue]{hyperref}

\begin{document}

\title{Frustration induced Itinerant Ferromagnetism of Fermions in Optical Lattices }

\author{Chengshu Li}
\affiliation{Institute for Advanced Study, Tsinghua University, Beijing 100084, China}
\author{Ming-Gen He}
\affiliation{Hefei National Research Center for Physical Sciences at the Microscale and School of Physics, University of Science and Technology of China, Hefei 230026, China}
\affiliation{CAS Center for Excellence in Quantum Information and Quantum Physics, University of Science and Technology of China, Hefei 230026, China}
\author{Chang-Yan Wang}
\affiliation{Institute for Advanced Study, Tsinghua University, Beijing 100084, China}
\author{Hui Zhai}
\email{hzhai@tsinghua.edu.cn}
\affiliation{Institute for Advanced Study, Tsinghua University, Beijing 100084, China}
\date{\today}

\begin{abstract}

When the Fermi Hubbard model was first introduced sixty years ago, one of the original motivations was to understand correlation effects in itinerant ferromagnetism. In the past two decades, ultracold Fermi gas in an optical lattice has been used to study the Fermi Hubbard model. However, the metallic ferromagnetic correlation was observed only in a recent experiment using frustrated lattices, and its underlying mechanism is not clear yet. In this letter, we point out that, under the particle--hole transformation, the single-particle ground state can exhibit double degeneracy in such a frustrated lattice. Therefore, the low-energy state exhibits valley degeneracy, reminiscent of multi-orbit physics in ferromagnetic transition metals. The local repulsive interaction leads to the valley Hund's rule, responsible for the observed ferromagnetism. We generalize this mechanism to distorted honeycomb lattices and square lattices with flux. This mechanism was first discussed by M\"uller-Hartmann in a simpler one-dimension model. However, this mechanism has not been widely discussed and has not been related to experimental observations before. Hence, our study not only explains the experimental findings but also enriches our understanding of itinerant ferromagnetism.  

\end{abstract}

\maketitle
Itinerant ferromagnetism has been discovered in nature for thousands of years. However, a complete understanding of its microscopic origin is still challenging. The Stoner mean-field theory has predicted itinerant ferromagnetism in metal when the repulsive interaction between fermions exceeds a critical value \cite{Stoner1938}. When such ferromagnetism occurs, the Pauli exclusion principle increases kinetic energy considerably. Therefore, the critical interaction strength predicted for Stoner ferromagnetism is comparable to the Fermi energy. Under such a strong interaction strength, the correlation effect can no longer be ignored. The competition from other strongly correlated non-magnetic states usually takes over Stoner ferromagnetism. Therefore, understanding itinerant ferromagnetism becomes essentially a strongly correlated problem.

When the Hubbard model was introduced in the middle of the last century \cite{Hubbard1963,Gutzwiller1963,Kanamori1963}, one major purpose was to understand the correlation effect in ferromagnetism. Unfortunately, the consensus on ferromagnetism in the Hubbard model is still limited after many decades \cite{Tasaki1998,Tasaki1998b,Arovas2022,Qin2022}. Rigorous results can only be obtained for exceptional cases such as the Nagaoka ferromagnetism \cite{Thouless1965,Nagaoka1966,Tasaki1989} and the flat band ferromagnetism \cite{Mielke1991}. The Nagaoka ferromagnetism considers a single hole doping away from half-filling in the limit of infinite repulsion, and the flat band ferromagnetism requires fine-tuning of particle hopping to reach a flat band dispersion. Moreover, the single-band Hubbard model is usually oversimplified to directly compare with experiments on real materials.  

Ultracold atoms in optical lattices directly realize the Hubbard model and provide a new opportunity to study physics therein \cite{Bloch2005,Bloch2008}. However, the temperature of atoms in optical lattices cannot be cooled much below the kinetic energy, which prevents observing possible low-temperature orderings, such as fermion pairing, at this stage. In contrast, ferromagnetism often occurs in a relatively high temperature. It is conceivable that one does not need to enter an extremely low-temperature regime in order to study ferromagnetism in optical lattices. Hence, understanding the nature of ferromagnetism is a potential goal for the current quantum simulations with optical lattices. 

Nevertheless, although the Fermi Hubbard model has been realized with ultracold atoms in optical lattices for nearly two decades, itinerant ferromagnetism has not been observed in this system until a very recent experiment~\cite{Xu2022}. In this experiment, a novel experimental technology allows continuously tuning the lattice geometry from a square to a frustrated triangular lattice. Short-range ferromagnetic correlation has been observed in the particle doping regime when the lattice geometry is tuned close to triangular regime. However, a convincing theoretical understanding of the physical mechanism behind the observed ferromagnetism is still lacking. 

\begin{figure}[t]
    \centering
    \includegraphics[width=0.45\textwidth]{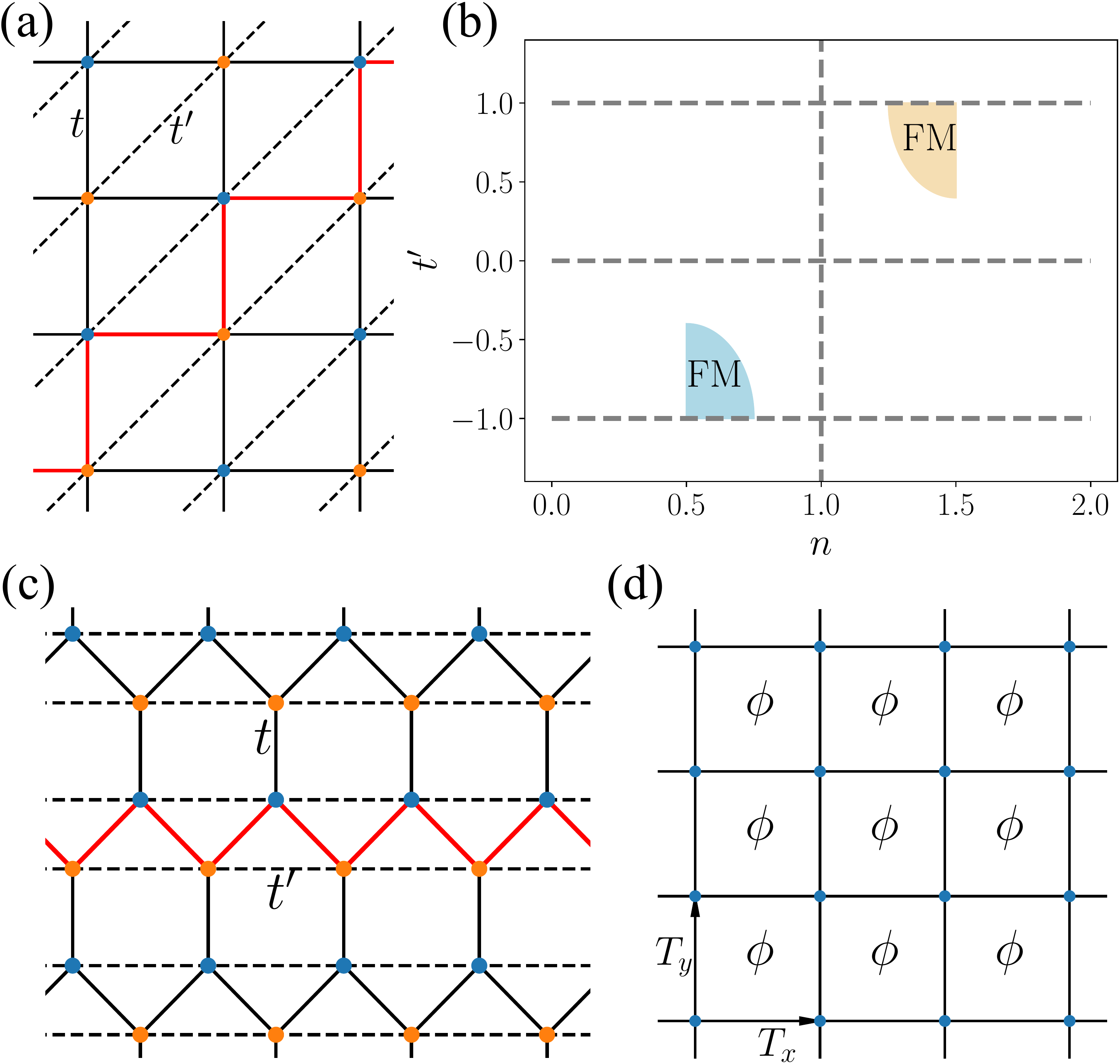}
    \caption{(a) The tunable optical lattice realized in a recent experiment. (b) The ferromagnetic correlation found in the experiment is marked by the shaded yellow regime in the $n-t^\prime$ phase diagram, which is mapped to the shaded blue area under the particle--hole transformation. (c) Honeycomb lattice with the nearest neighbor and the horizontal next nearest neighbor hopping. (d) Square lattice with the nearest neighbor hopping only, but with magnetic flux in each plaquette. The red lines in (a) and (c) denote the one-dimensional chain considered by M\"uller-Hartmann's paper.}
     \label{model}
\end{figure}

\textit{Review of Experimental Setting.} This experiment reports an actively phase stabilized optical lattice that can continuously tune the lattice geometry from square lattice to triangular lattice. The tight-binding model of this lattice is shown in Fig.~\ref{model}(a). This experiment explores spin-$1/2$ fermions in such lattice, and the model is written as
\begin{equation}
\hat{H}=-t\sum\limits_{\langle ij\rangle\sigma}\hat{c}^\dag_{i\sigma}\hat{c}_{j\sigma}-t^\prime \sum\limits_{\langle\langle ij \neswarrow\rangle\rangle}\hat{c}^\dag_{i\sigma}\hat{c}_{j\sigma}+U\sum\limits_{i}\hat{n}_{i\uparrow}\hat{n}_{i\downarrow}, \label{Hubbard}
\end{equation}
where $\sigma=\uparrow,\downarrow$ denotes two spin components. The hopping between the nearest neighbor $\langle ij\rangle$ is denoted by $t$, and between the next nearest neighbor along the dashed line direction $\langle\langle ij \neswarrow\rangle\rangle$ is denoted by $t^\prime$. Both $t$ and $t^\prime$ are positive. The next nearest hopping along another diagonal direction is negligible. When $t^\prime$ increases from zero to $t^\prime=t$, it continuously tunes the lattice geometry from a square lattice to a triangular lattice. $U$ denotes the interaction strength of on-site repulsion, and $U/t \approx 9$ in this experiment. 

This experiment finds ferromagnetic correlation when $t^\prime/t>0.5$ and when total fermion density $n$ exceeds half-filling $n=1$ and is somewhat close to $n=1.5$. The regime where ferromagnetism is observed is marked by shaded yellow area in Fig.~\ref{model}(b). This experiment has only explored the density regime $0.5<n<1.5$ and it is not clear whether the ferromagnetic correlation can also exist when $n$ exceeds $1.5$. The physical origin of this observed ferromagnetism is also not clear yet. Possible scenarios mentioned include the existence of van Hove singularity at $n=1.5$ for triangular lattice and the possible connection to the Nagaoka ferromagnetism \cite{Xu2022,Merino2006,Demler2022}.   

\textit{Particle--Hole Transformation.} For the benefit of later discussion, we make a particle--hole transformation $\hat{c}_{i\sigma}\rightarrow (-1)^{i_x+i_y}\hat{c}^\dag_{i\sigma}$, where $i=(i_x,i_y)$ is the site label. If $t^\prime=0$, this transformation keeps the form of Hamiltonian Eq.~\eqref{Hubbard} invariant. $t^\prime$-term is what causes frustration in this lattice, and for the same reason, $t^\prime$-term is not invariant under the particle--hole transformation. Instead, this transformation changes $t^\prime$ to $-t^\prime$. This corresponds to inserting a $\pi$-flux in each triangle. For positive $t^\prime$, the system essentially describes particles moving in a real potential and all hopping terms have negative matrix elements that satisfies Feynman's no-node theorem~\cite{Feynman1998}. Therefore, the single-particle ground state cannot have degeneracy. However, for negative $t^\prime$, the condition for no-node theorem is no longer satisfied, and the single-particle ground state can have degeneracy. It turns out that this degeneracy is crucial for explaining this observed ferromagnetism, as it will become clear later. 

Under this transformation, $n=n_\uparrow+n_\downarrow$ becomes $2-n$, that is, the particle doping is mapped to hole doping. Thus, the particle--hole transformation maps $(t^\prime,n)$ to $(-t^\prime,2-n)$ in the phase diagram shown in Fig.~\ref{model}(b). Therefore, we can focus solely on the density regime $0<n\leqslant 1$ but include both positive and negative $t^\prime$. With this mapping, the regime where ferromagnetism is observed is mapped to the low-density regime with $t^\prime<-0.5$, as marked as shaded blue area in Fig.~\ref{model}(b). 

\begin{figure}[t]
    \centering
    \includegraphics[width=0.45\textwidth]{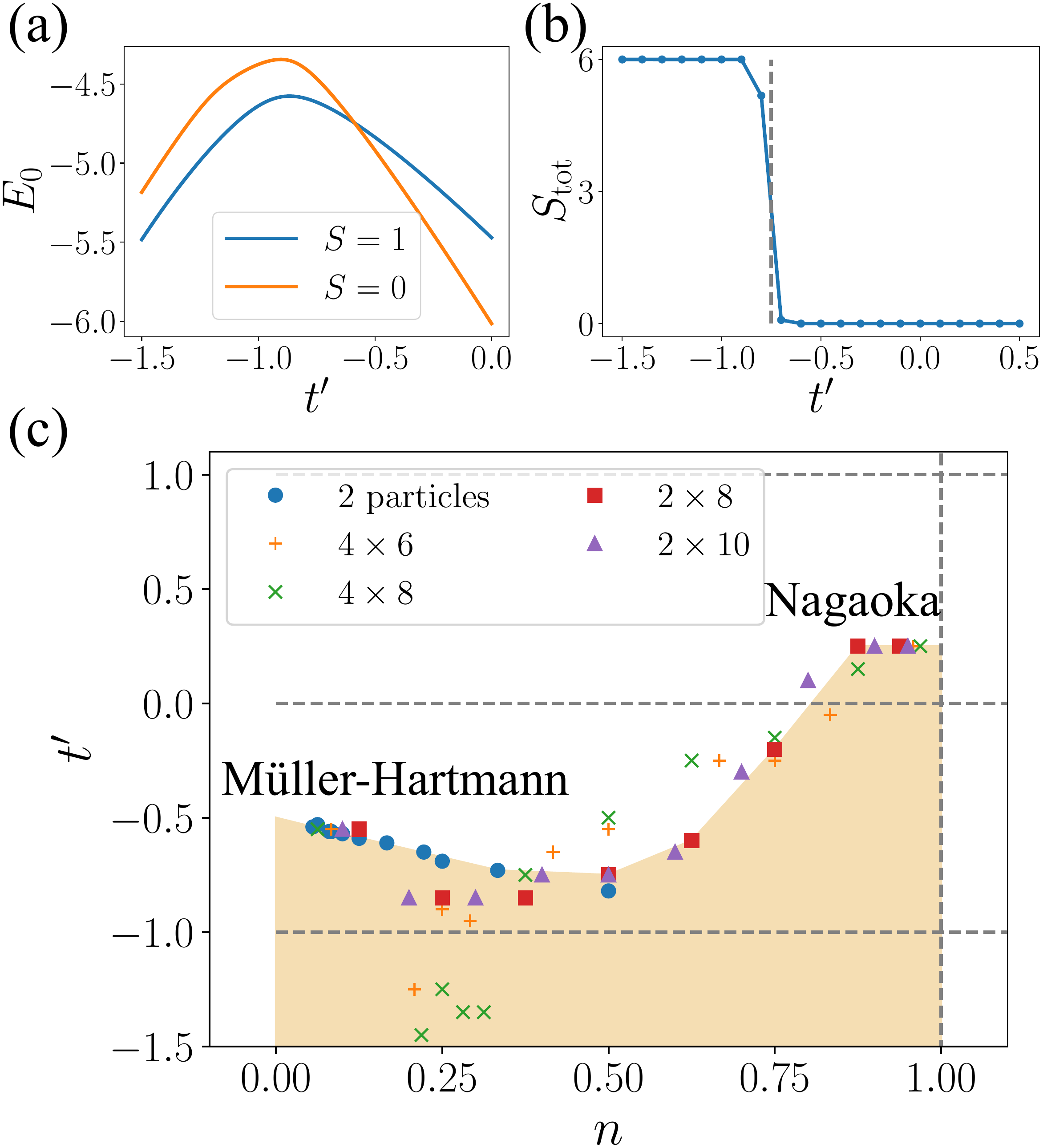}
    \caption{(a) Two-body problem in a $4\times 4$ lattice with open boundary condition.  The singlet and triplet state energies as a function of $t^\prime$. (b) $S_\text{tot}$ of the ground state calculated by the DMRG method for $N=12$ particles in a $4\times 8$ strip. (c)~The shaded area schematically shows the possible ferromagnetic regime in the $n-t^\prime$ phase diagram. Each point denote $t^\prime_\text{c}$ between ferromagnetic and non-magnetic phase at different  densities. The numerical calculations include exact diagonalizations of two-particle problem in different system sizes and DMRG calculations of finite number of particles on different system sizes. All DMRG calculations have bond dimension $1400$. In all cases, $t=1$ and $U$ is set to positive infinite. }
     \label{Square}
\end{figure}

\textit{Numerical Results.} We first present our numerical results in Fig.~\ref{Square} \cite{code}. All the calculations below are done for infinite positive $U$. The first calculation is exact diagonalization for two particles with different system sizes, reminiscent of different densities. This calculation can cover the low-density regime up to $n=0.5$.  Due to the $SU(2)$ spin rotational symmetry, the total spin is a good quantum number and all quantum states with the same total spin are degenerate. We find a level crossing between spin singlet and spin triplet states, as shown in Fig.~\ref{Square}(a). The spin triplet states have a lower energy when $t^\prime<t^\prime_\text{c}$, and the value of $t^\prime_\text{c}$ depends on density and lies between $-1.0$ and $-0.5$, as one can see from circles in Fig.~\ref{Square}(c). 

The second calculation is the density-matrix renormalization group (DMRG) calculation with finite number of fermions on different strip geometry. We can calculate $\langle \hat{{\bf S}}_\text{tot}^2\rangle$ with $\hat{{\bf S}}_\text{tot}=\sum_{i}\hat{{\bf S}}_i$ for the ground state, and $S_\text{tot}$ is given by $\langle \hat{{\bf S}}_\text{tot}^2\rangle=S_\text{tot}(S_\text{tot}+1)$. We also find a transition from $S_\text{tot}=N/2$ to $S_\text{tot}=0$ around $t^\prime=t^\prime_\text{c}$, as shown in Fig.~\ref{Square}(b). Fig.~\ref{Square}(c) also collects $t^\prime_\text{c}$ obtained by all DMRG calculations with different fermion number and different system sizes. 

A notable feature in Fig.~\ref{Square}(c) is that for all calculations, $t^\prime_\text{c}$ approaches $-0.5$ at the low-density limit when $n\rightarrow 0$. Below we will explain why $t^\prime=-0.5$ is a special point, and the ferromagnetism emerged in this regime will be called the \textit{M\"uller-Hartmann Mechanism}. In this regime, the general trend is that $t^\prime_\text{c}$ decreases as $n$ increases. 

\begin{figure}[t]
    \centering
    \includegraphics[width=0.45\textwidth]{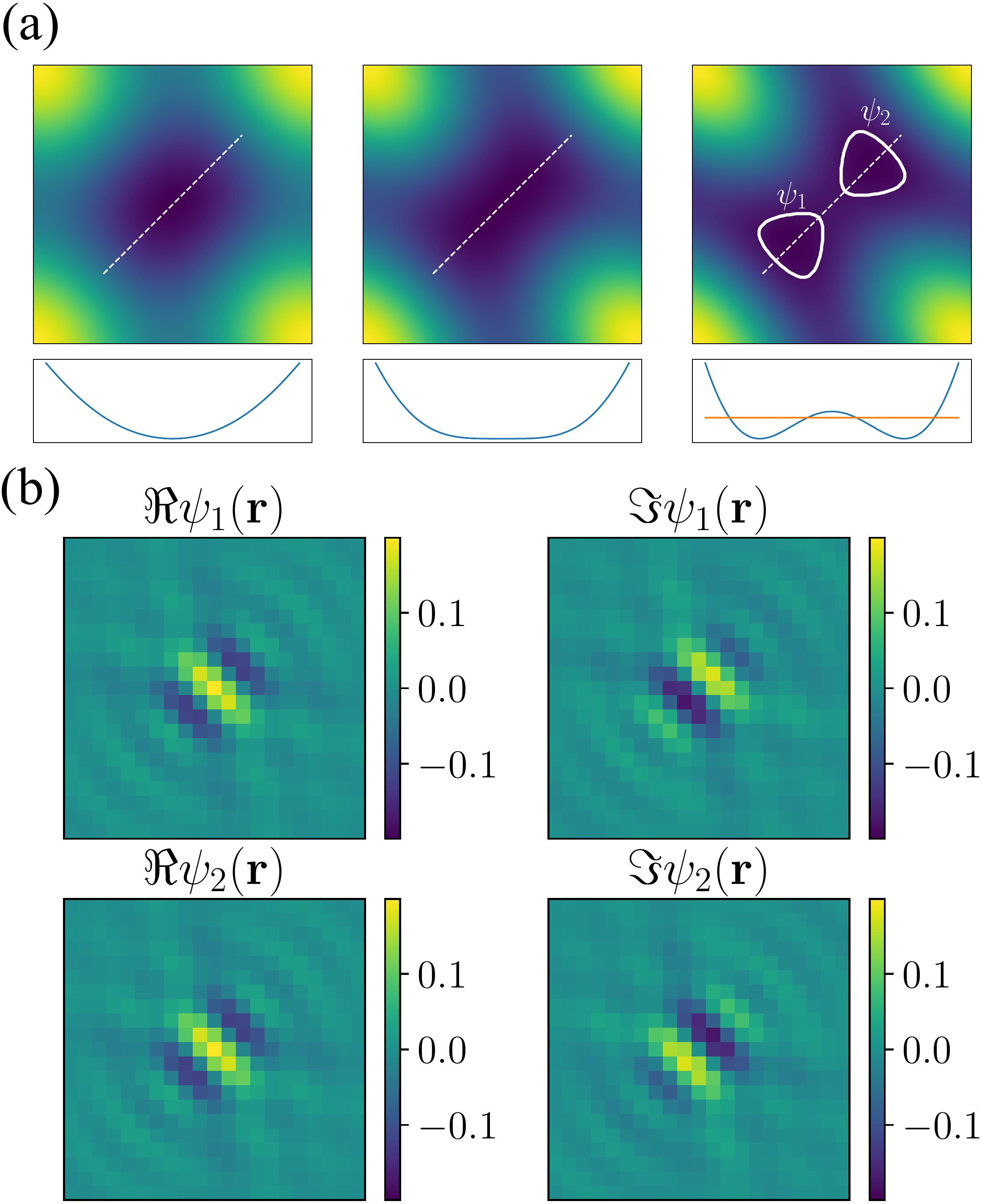}
    \caption{(a) Single-particle dispersion $\mathcal{E}(k_x,k_y)$ of the Hamiltonian Eq.~\eqref{Hubbard} for three different values of $t^\prime$. $t^\prime=-0.2$ (left), $=-0.5$ (middle) and $=-0.8$ (right). (b) The real and imaginary parts of two valley Wannier wave functions. These two wave functions are constructed by using the Bloch wave functions around each minimum of single-particle dispersion, as indicated by the solid circle in the right dispersion in (a).  }
     \label{wannier}
\end{figure}

Nearby half-filling with $n=1$, it is known that hole doping can result in Nagaoka ferromagnetism at infinite $U$. Strictly speaking, Nagaoka ferromagnetism can only be proved for single-hole doping with $t^\prime<0$. However, our numerical results show that when $t^\prime=0$, the Nagaoka ferromagnetism can exist up to $\sim 20\%$ of hole doping, that is, for $0.8\lesssim n<1$. This is consistent with previous numerical results~\cite{Liu2012}. Hence, we attribute the ferromagnetism nearby half-filling as \textit{Nagaoka Mechanism}. In this regime, $t^\prime_\text{c}$ also decreases when $n$ decreases from $n=1$, and this  density dependence of $t^\prime_\text{c}$ is opposite to that found in the low-density regime. 

Hence, we have identified two different mechanisms of ferromagnetism from our numerical results, and they emerge from the low-density limit and nearby half-filling, respectively. They are characterized by different density dependence of $t^\prime_\text{c}$. The trend of experimental observed ferromagnetic boundary, upon the particle--hole transformation, is consistent with the M\"uller-Hartmann mechanism. 

\begin{figure}[t]
    \centering
    \includegraphics[width=0.45\textwidth]{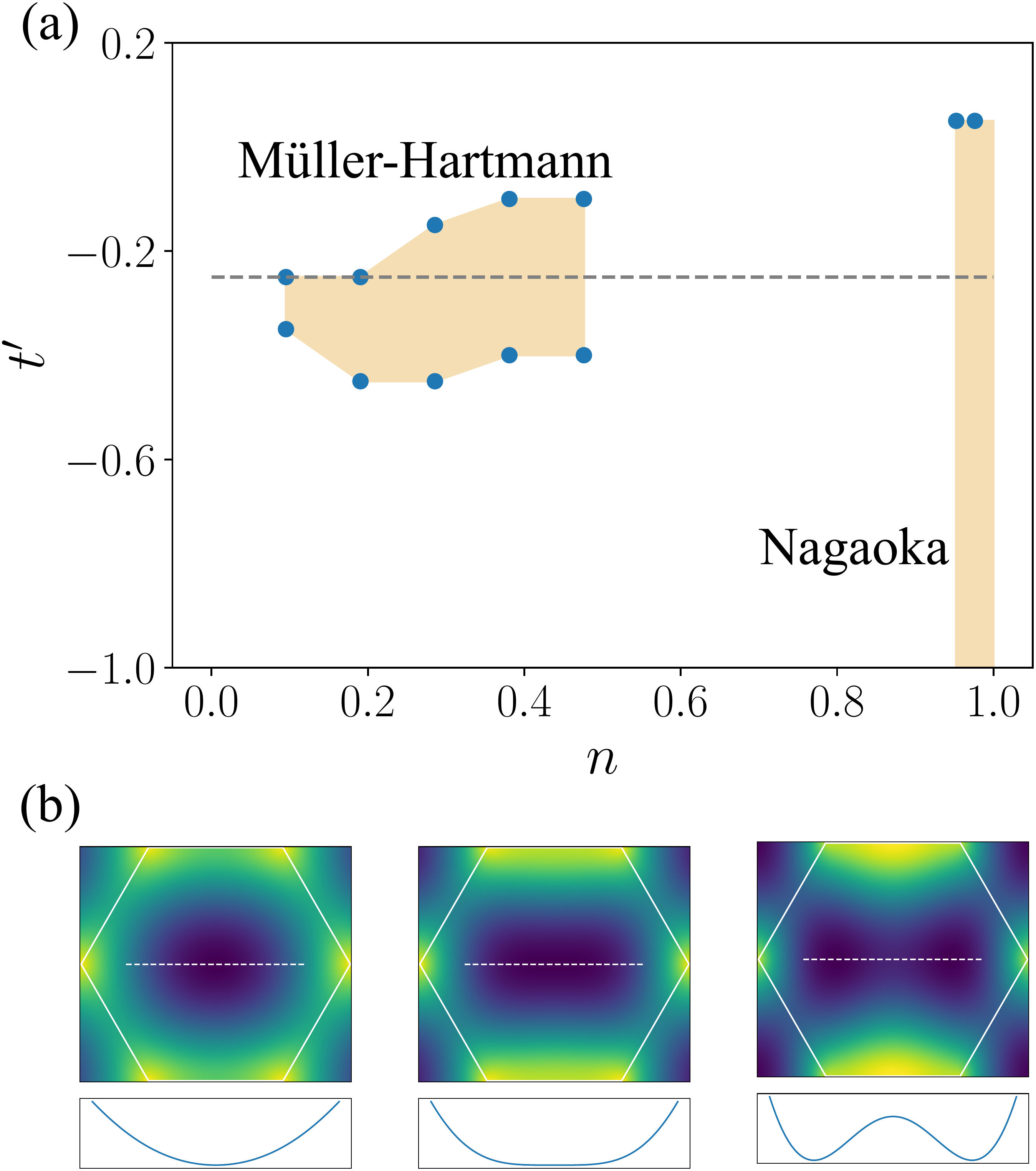}
    \caption{(a) The ferromagnetic regime indicated by DMRG calculation in the $n-t^\prime$ phase diagram of the distorted honeycomb lattice. The model is shown in Fig.~\ref{model}(c), in which $t^\prime$ denotes the next nearest hopping along the horizontal dashed line. The DMRG calculation is performed in a $3\times 6$ zigzag cylinder, with bond dimension $\chi=1400$. The dashed line indicates the value of $t^\prime$ for the Lifshitz transition. (b) The single-particle dispersion of the distorted honeycomb lattice, with $t^\prime=-0.1$ (left), $=-0.25$ (middle) and $=-0.4$ (right). }
     \label{Honeycomb}
\end{figure}

\textit{M\"uller-Hartmann Mechanism.} To see why $t^\prime=-0.5$ is special, we first look at the single-particle dispersion which reads
\begin{align}
\mathcal{E}(k_x,k_y)&=-2t\cos(k_x)-2t\cos(k_y)-2t^\prime\cos(k_x+k_y)\nonumber\\
&=-4t\cos(k_+)\cos(k_{-})-2t^\prime\cos(2k_{+}),
\end{align}
where $k_{\pm}=(k_x\pm k_y)/2$. It is easy to see that the dispersion minimum occurs at $k_{-}=0$. As shown in Fig.~\ref{wannier}(a), when $t^\prime/t>-0.5$, the band dispersion has a unique minimum at $k_{+}=0$. However, when $t^\prime/t<-0.5$, the band dispersion displays two degenerate minima along $k_{+}$ axes. That is to say, a Lifshitz transition occurs when $t^\prime/t=-0.5$.

The paper by M\"uller-Hartmann first pointed out that ferromagnetism can occur in the low-density regime when the band minima display double degeneracy \cite{Muller1995}, followed by few related works in later literatures \cite{Pieri1996}. This paper considered a one-dimensional chain with next nearest hopping, and the Hamiltonian reads
\begin{equation}
\hat{H}=-\sum\limits_{i\sigma}(t\hat{c}^\dag_{i,\sigma}\hat{c}_{i+1,\sigma}+t^\prime\hat{c}^\dag_{i,\sigma}\hat{c}_{i+2,\sigma}+\text{h.c.})+U\sum\limits_{i}\hat{n}_{i\uparrow}\hat{n}_{i\downarrow}.
\end{equation}
In this model, the band minima also displays double degeneracy when $t^\prime/t<-1/4$, where M\"uller-Hartmann argued that ferromagnetism can occur in the low-density regime. We note that this one-dimensional chain can be naturally embedded into the two-dimensional lattice we considered, as shown in Fig.~\ref{model}(a). In other word, the two-dimensional lattice realized in this cold atom experiment can be viewed as a two-dimensional generalization of M\"uller-Hartmann's work.

The physical mechanism behind this ferromagnetism can be called the \textit{Valley Hund's rule} \cite{valley}. Considering the two-particle case, they can be respectively placed in each minimum of single-particle dispersion. For spin-triplet states, the spatial wave function is anti-symmetrized, and the interaction energy automatically vanishes. However, for the spin-singlet state, one has to apply extra projection to prevent double occupation, which inevitably increases the kinetic energy.

Using the low-energy Bloch wave functions around each minimum, we can construct the effective valley Wannier wave functions for each valley, as shown in Fig.~\ref{wannier}(b). These valley Wannier wave functions extend over several lattice sites. The on-site repulsion between fermions effectively introduces Hund's rule between these valley Wannier wave functions. In real materials, itinerant ferromagnetism usually occurs in transition metals with partially filled $d$-orbit, and Hund's rule coupling between these $d$-orbits is crucial for ferromagnetism. That is to say, the multi-orbital physics plays an essential role. M\"uller-Hartmann's mechanism says that even for a single-band model, when the single-particle ground state has degeneracy, the valley degeneracy can lead to an emergent multi-orbital physics, resulting in itinerant ferromagnetism at the low-density regime. Since Hund's rule is a local ferromagnetic coupling, we expect that the short-range ferromagnetic correlation caused by this mechanism is robust at finite temperatures. Moreover, since the Hund's rule coupling is the leading order effect of local repulsion while the anti-ferromagnetic super-exchange interaction is a second-order effect of local repulsion, we also expect that this ferromagnetism is stable with finite interaction strength.     

\begin{figure}[t]
    \centering
    \includegraphics[width=0.45\textwidth]{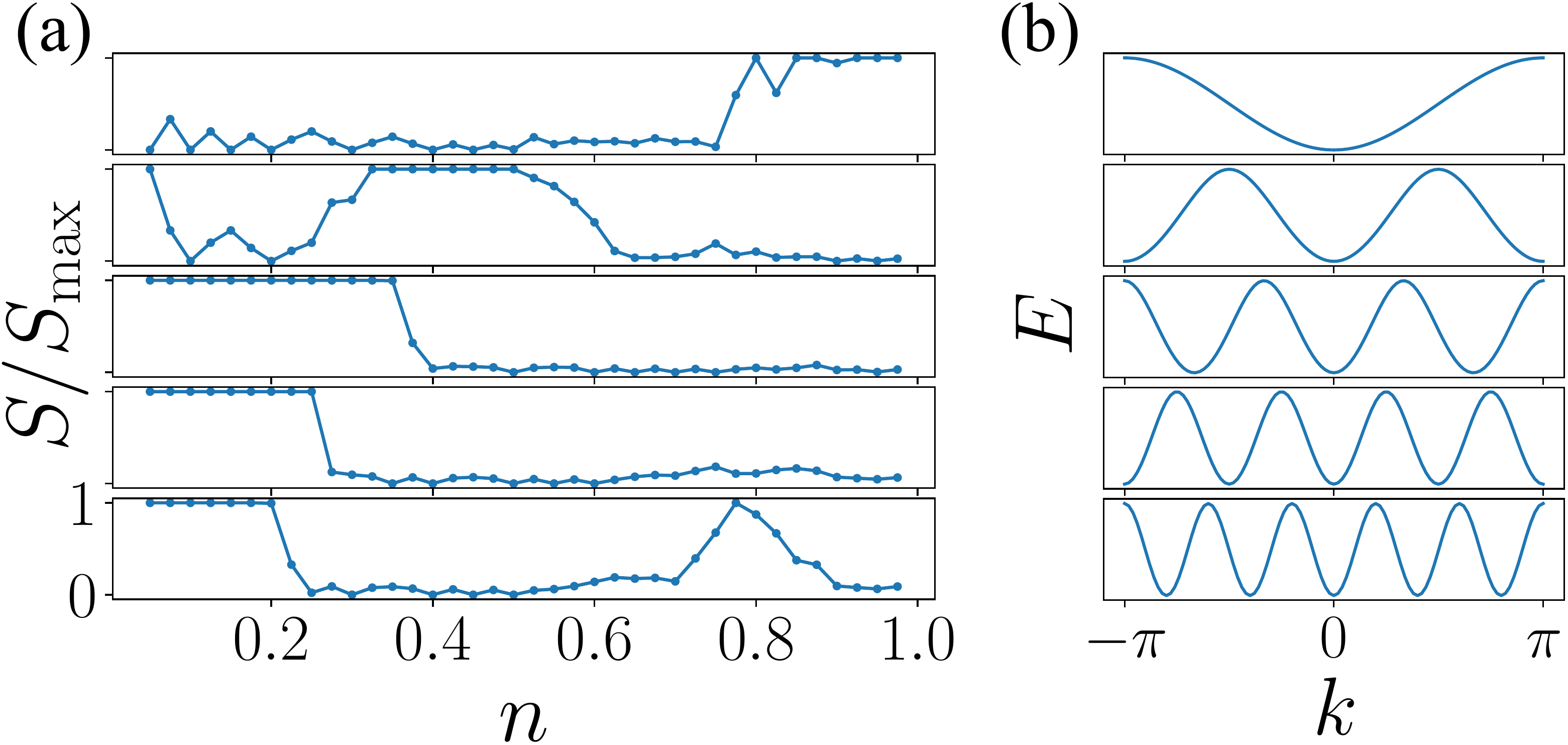}
    \caption{(a) $S_\text{tot}$ obtained by DMRG calculation. $S_\text{max}=N/2$. Here we consider square lattice with nearest hopping only, but with magnetic flux $\phi$. $\phi=0,\pi,2\pi/3,\pi/2$ and $2\pi/5$ from the top to the bottom. The DMRG calculation is performed in a $4\times 10$ strip with bond dimension $\chi=1400$. (b) The single-particle dispersion around the bottom of the band for different magnetic flux corresponding to (a).  }
     \label{Flux}
\end{figure}

\textit{Generalization to Honeycomb Lattice.} To further confirm our understanding of emergent ferromagnetism in this experiment, we generalize this idea to other models, where degenerate ground states can also be found. The first generalization is fermions in a distorted honeycomb lattice. Similar to the square lattice model, the nearest neighbor hopping is denoted by $t$. We compress the lattice along the horizontal direction such that we should also include the next nearest hopping $t^\prime$ along the horizontal dash line direction, as shown Fig.~\ref{model}(c). This lattice can also be viewed as a set of $t-t^\prime$ chains along $\hat{x}$ direction considered by M\"uller-Hartmann's paper, coupled vertically along $\hat{y}$ direction. We introduce a similar particle-hole transformation that only changes sign of $t^\prime$. The ground state of this dispersion also shows a Lifshitz transformation from single to double degeneracy at $t^\prime=-0.25$, as shown in Fig.~\ref{Honeycomb}(b). The phase diagram obtained by DMRG calculation is shown in Fig.~\ref{Honeycomb}(a). This phase diagram contains two ferromagnetic regimes, one in the low-density regime around $t^\prime=-0.25$ and the other nearby half-filling for $t^\prime<0$. Clearly, they respectively manifest M\"uller-Hartmann and Nagaoka mechanisms \cite{Bobrow2018}. Unlike the square lattice case, these two ferromagnetism mechanisms occur in two disconnected regimes in the phase diagram, clearly visualizing their difference.     

\textit{Generalization to Flux Lattice.} The second generalization is square lattice with a magnetic flux in each plaquette. Here we only consider the nearest hopping in square lattice, and the magnetic flux only introduces a phase in the hopping and does not cause Zeeman splitting. However, with magnetic flux $\phi=2\pi/m$ in each plaquette ($m$ is an integer), translation of one lattice spacing along $\hat{x}$ direction does not commute with one lattice space translation along $\hat{y}$ direction, and only commutes with translation of $m$ lattice spacing along $\hat{y}$ direction. This enlarges the unit cell and leads to $m$-fold degeneracy of dispersion spectrum, as shown in Fig.~\ref{Flux}(b). For a fixed magnetic flux, the only tunable parameter is density. In Fig.~\ref{Flux}(a), we show the total magnetization as a function of fermion density. It is interesting to note that when $m=1$, the single-particle dispersion is not be degenerate but the hopping satisfies the condition for Nagaoka ferromagnetism. Hence, DMRG finds ferromagnetism nearby half-filling. In contrast, when $m>1$, hopping terms no longer satisfy the condition for Nagaoka ferromagnetism, but the ground state degeneracy appears. We find a critical $n_\text{c}$ and the system is ferromagnetism when $n<n_\text{c}$. We find that $n_\text{c}$ is approximately given by $1/m$, which corresponds to a half-filled lowest band. Therefore, ferromagnetism occurs in low-density regime  \cite{m2}. This clearly shows Nagaoka and M\"uller-Hartmann are two different mechanisms.  

\textit{Conclusion and Discussions.} In summary, we analyze a recent experiment that first discovers itinerant ferromagnetism in ultracold atom realization of the Fermi Hubbard model. We attribute the mechanism of this ferromagnetism to the appearance of single-particle ground-state degeneracy. At low density, the valley degeneracy is reminiscent of multi-orbital physics in transition metals, and the local repulsive interaction can lead to a valley Hund's rule, resulting in ferromagnetic correlations. This mechanism was first discussed by M\"uller-Hartmann in a simpler one-dimension model. 

We remark that by talking about ferromagnetism, we implicitly assume that the Hamiltonian of the system should not break time-reversal symmetry. In the presence of time-reversal symmetry, the single-particle ground state should not degenerate for normal kinetic energy term due to Feynman's no-node theorem. In order for the M\"uller-Hartmann mechanism to apply, the key is frustration plus particle-hole transformation. Because of the lattice frustration, we can effectively inset flux to the hopping terms after a particle--hole transformation, such that the condition for Feynman's theorem is violated. That is why frustration plays an essential role here. Previously, frustration has been considered an essential ingredient for the emergence of exotic phases such as spin liquids~\cite{Balents2010}; the discussion here shows that frustration can also be a key ingredient responsible for the emergence of ferromagnetism. 

\textit{Acknowledgement}. We thank Hong Yao and Yingfei Gu for helpful discussions. This work is supported by Innovation Program for Quantum Science and Technology 2021ZD0302005, the Beijing Outstanding Young Scholar Program, the XPLORER Prize and China Postdoctoral Science Foundation (Grant No. 2022M711868). C.L. is supported by Chinese International Postdoctoral Exchange Fellowship Program and Shuimu Tsinghua Scholar Program at Tsinghua University.


\begin{thebibliography}{99}
\bibitem{m2}
Here for $\phi=\pi$ case, the non-magnetic state at $n\sim 0.1-0.3$ is anomalous. We find the results in this density range with $\phi=\pi$ quite sensitive to system geometry, and the finite size effect is quite significant in this density range.
\end{thebibliography}
\end{document}